\documentclass{article}
\usepackage{spconfa4,amsmath,graphicx}
\usepackage{amsmath,graphicx,url,times,multirow}
\usepackage{color}
\usepackage{cite}
\usepackage{url}

\newcommand{\etal}{\textit{et al.}}
\newcommand{\cmmnt}[1]{\ignorespaces}

\title{MMDenseLSTM: An efficient combination of convolutional and recurrent neural networks for audio source separation}
\name{Naoya Takahashi$^1$, Nabarun Goswami$^2$, Yuki Mitsufuji$^1$}
\address{$^1$Sony Corporation, Minato-ku, Tokyo, Japan\\
$^2$Sony India Software Centre, Bangalore, India}
\begin{document}
%
\maketitle
\begin{abstract}
Deep neural networks have become an indispensable technique for audio source separation (ASS). It was recently reported that a variant of CNN architecture called MMDenseNet was successfully employed to solve the ASS problem of estimating source amplitudes, and state-of-the-art results were obtained for DSD100 dataset. To further enhance MMDenseNet, here we propose a novel architecture that integrates long short-term memory (LSTM) in multiple scales with skip connections to efficiently model long-term structures within an audio context. The experimental results show that the proposed method outperforms MMDenseNet, LSTM and a blend of the two networks. The number of parameters and processing time of the proposed model are significantly less than those for simple blending. Furthermore, the proposed method yields better results than those obtained using ideal binary masks for a singing voice separation task.
\end{abstract}

\section{Introduction}
\label{sec:intro}


Audio source separation (ASS) has recently been intensively studied. Various approaches have been introduced such as local Gaussian modeling \cite{DuongVG10, FitzgeraldLB16}, non-negative factorization \cite{LiutkusFB15, Roux15, MitsufujiKS16}, kernel additive modeling \cite{LiutkusFRPD14} and their combinations \cite{OzerovF10, LiutkusFR15, FitzgeraldLB162}. 
Recently, deep neural network (DNN) based ASS methods have shown a significant improvement
over conventional methods. 
In \cite{Nugraha15, Uhlich15}, a standard feedforward fully connected network (FNN) was used to estimate source spectra. A common way to exploit temporal contexts is to concatenate multiple frames as the input. However, the number of frames that can be used is limited in practice to avoid the explosion of the model size.
In \cite{Uhlich17}, long short-term memory (LSTM), a type of recurrent neural network (RNN), was used to model longer contexts. However, the model size tends to become excessively large and the training becomes slow owing to the full connection between the layers and the gate mechanism in an LSTM cell.
Recently, convolutional neural networks (CNNs)~\cmmnt{~\cite{LeCun1998}} have been successfully applied to audio modeling of spectrograms \cite{Sercu2015, Takahashi2016, Korzeniowski16, Takahashi2017AENet}, although CNNs were originally introduced to address the transition-invariant property of images. A CNN significantly reduces the number of parameters and improves generalization by sharing parameters to model local patterns in the input.  However, a standard CNN requires considerable depth to cover long contexts, making training difficult. To address this problem, a multi-scale structure was used to adapt a CNN to solve the ASS problem in \cite{Jansson17, Takahashi17MMDense}, where convolutional layers were applied on multiple {\it scales} obtained by downsampling feature maps, and low-resolution feature maps were progressively upsampled to recover the original resolution. Another problem in applying a two dimensional convolution to a spectrogram is the biased distribution of the local structure in the spectrogram. Unlike an image, a spectrogram has different local structures depending on the frequency bands. Complete sharing of convolutional kernels over the entire frequency range may reduce modeling flexibility. In \cite{Takahashi17MMDense}, we proposed a multi-band structure where each band was linked to a single CNN dedicated to particular frequency bands \cmmnt{which combined} along with a full-band CNN. The novel CNN architecture called DenseNet was extended to the multi-scale and multi-band structure called MMDenseNet, which outperformed an \cmmnt{bi-directional} LSTM system and achieved a state-of-the-art performance for the DSD100 dataset \cite{Liutkus17}.
Although it has been suggested that CNNs often work better than RNNs even for sequential data modeling \cite{Bai18, Takahashi17MMDense}, RNNs can also benefit from multi-scale and multi-band modeling because
they make it easier to capture long-term dependencies and can save parameters by omitting redundant connections between different bands.

Moreover, the blending of two systems improves the performance even when one system consistently performs better than the other\cite{Uhlich17}. Motivated by these observations, here we propose a novel network architecture called MMDenseLSTM. This combines LSTM and DenseNet in multiple scales and multiple bands, improving separation quality while retaining a small model size. 
There have been several attempts to combine CNN and RNN architectures. In \cite{Sainath15, Zhao18}, convolutional layers and LSTM layers were connected serially \cmmnt{and applied to automatic speech recognition}. Shi \etal~proposed convolutional LSTM for the spatio-temporal sequence modeling of rainfall prediction \cite{Shi15}, where matrix multiplications in LSTM cells were replaced with convolutions. In contrast to these methods, in which convolution and LSTM operate at a single scale, we show that combining them at multiple low scales increases the performance and efficiency. Moreover, we systematically compare several architectures to search for the optimal strategy to combine DenseNet and LSTM. Experimental results show that the proposed method outperforms current state-of-the-art methods for the DSD100 and MUSDB18 datasets. Furthermore, MMDenseLSTM even outperforms an ideal binary mask (IBM), which is usually considered as an upper baseline, when we train the networks with a larger dataset. 

\section{Multi-scale multi-band DenseLSTM}
In this section, we first summarize multi-scale multi-band DenseNet (MMDenseNet) as our base network architecture. Then, we introduce strategies to combine {\it dense block}  and LSTM at multiple scales and multiple bands. 
Finally, we discuss the architecture of MMDenseLSTM in detail.

\subsection{MMDenseNet}
\label{sec:MMDenseNet}
\begin{figure}[t]
\centering
\includegraphics[width=\linewidth]{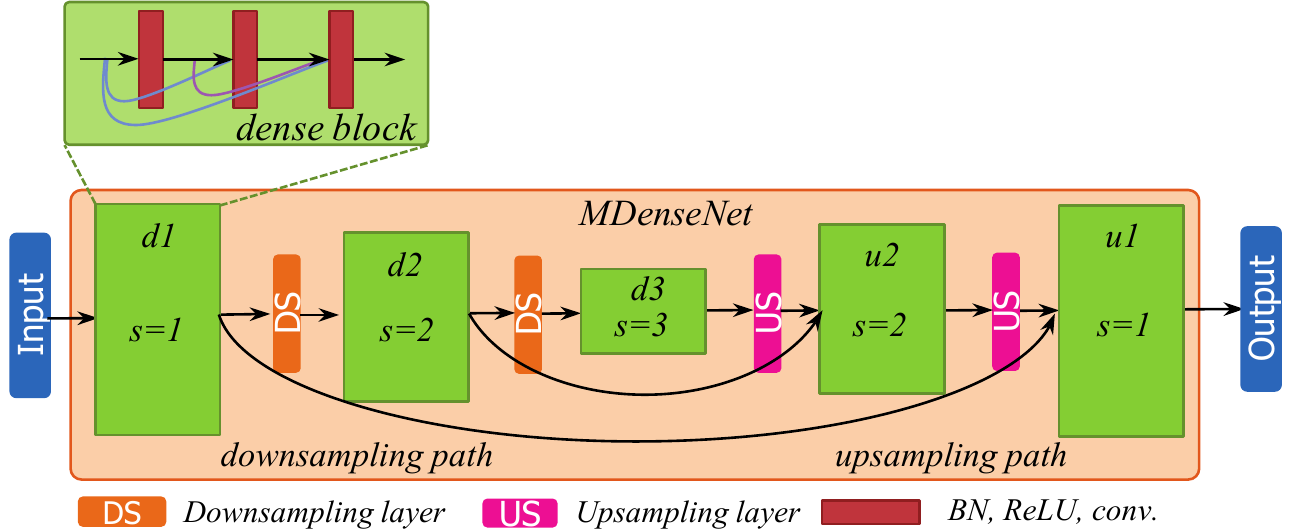}
\caption{MDenseNet architecture. Multi-scale dense blocks are connected though down- or upsampling layer or through block skip connections. The figure shows the case $s=3$.}
\label{fig:mdense}
\end{figure}
Among the various CNN architectures, DenseNet shows excellent performance in image recognition tasks \cite{Huang2016}. The basic idea of DenseNet is to improve the information flow between layers by concatenating all preceding layers as,
$x_l = H_l([x_{l-1}, x_{l-2},\ldots,x_0])$,
where $x_l$ and $[\ldots]$ denote the output of the $l$th layer and the concatenation operation, respectively.  $H_l(\cdot)$ is a nonlinear transformation consisting of batch normalization (BN) followed by ReLU and convolution with $k$ feature maps. Such dense connectivity enables all layers to receive the gradient directly and also reuse features computed in preceding layers. To cover the long context required for ASS, multi-scale DenseNet (MDenseNet) applies a dense block at multiple scales by progressively downsampling its output and then progressively upsampling the output to recover the original resolution, as shown in Fig.\ref{fig:mdense}. Here, $s$ is the scale index, i.e., the feature maps at scale $s$ are downsampled $s-1$ times and have $2^{2(s-1)}$ times lower resolution than the original feature maps. To allow forward and backward signal flow without passing though lower-resolution blocks, an inter-block skip connection, which directly connects two dense blocks of the same scale, is also introduced. 

In contrast to an image, in an audio spectrogram, different patterns occur in different frequency bands, although a certain amount of translation of patterns exists for a relatively small pitch shift. Therefore, limiting the band that shares the kernels is suitable for efficiently capturing local patterns. MMDenseNet addresses this problem by splitting the input into multiple bands and applying band-dedicated MDenseNet to each band.  
MMDenseNet has demonstrated state-of-the-art performance for the DSD100 dataset with about 16 times fewer parameters than the LSTM model, which obtained the best score in SiSEC 2016 \cite{Liutkus17}. 

\label{sec:MMDenseNet}

\subsection{Combining LSTM with MMDenseNet}
\begin{figure}[t]
\centering
\includegraphics[width=\linewidth]{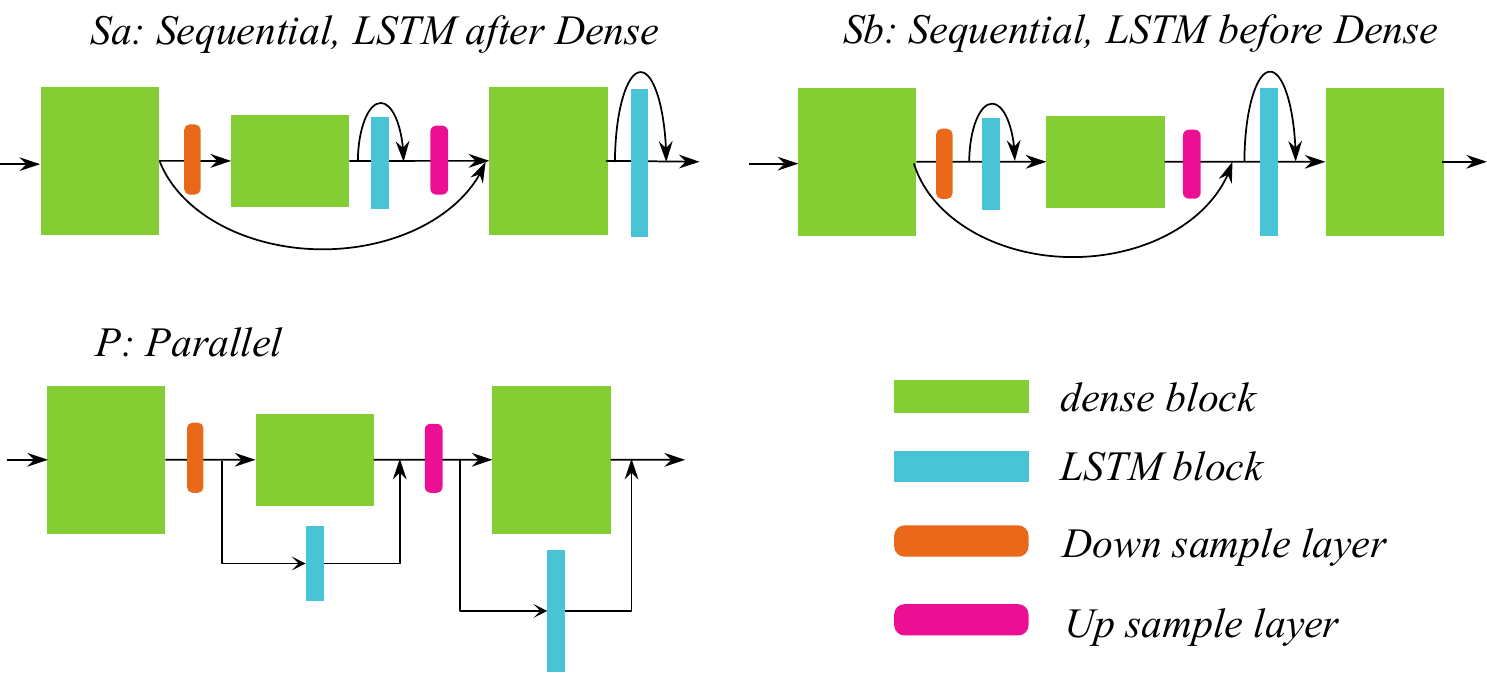}
\caption{Configurations with different combinations of dense and LSTM blocks. LSTM blocks are inserted at some of the scales}
\label{fig:rmdensetypes}
\end{figure}

Uhlich \etal~have shown that blending two systems gives better performance even when one system consistently outperforms the other \cite{Uhlich17}. The improvement tends to be more significant when two very different architectures are blended such as a CNN and RNN, rather than the same architectures with different parameters. However, the blending of architectures increases the  model size and computational cost in an additive manner, which is often undesirable when deploying the systems. Therefore, we propose combining the dense block and {\it LSTM block} in a unified architecture. The {\it LSTM block} consists of a $1\times1$ convolution that reduces the number of feature maps to $1$, followed by a bi-directional LSTM layer, which treats the feature map as sequential data along the time axis, and finally a feedforward linear layer that transforms back the input frequency dimension $f^s$ from the number of LSTM units $m^s$. We consider three configurations with different combinations of the dense and LSTM blocks as shown in Fig. \ref{fig:rmdensetypes}. The {\it Sa} and {\it Sb} configurations place the LSTM block after and before the dense block, respectively, while the dense block and LSTM block are placed in parallel and concatenated in the {\it P} configuration. We focus on the use of the {\it Sa} configuration since a CNN is effective at modeling the local structure and the LSTM block benefits from local pattern modeling as it covers the entire frequency at once. This claim will be empirically validated in Sec. \ref{sec:arcvalid}.

\begin{figure}[t]
\centering
\includegraphics[width=\linewidth]{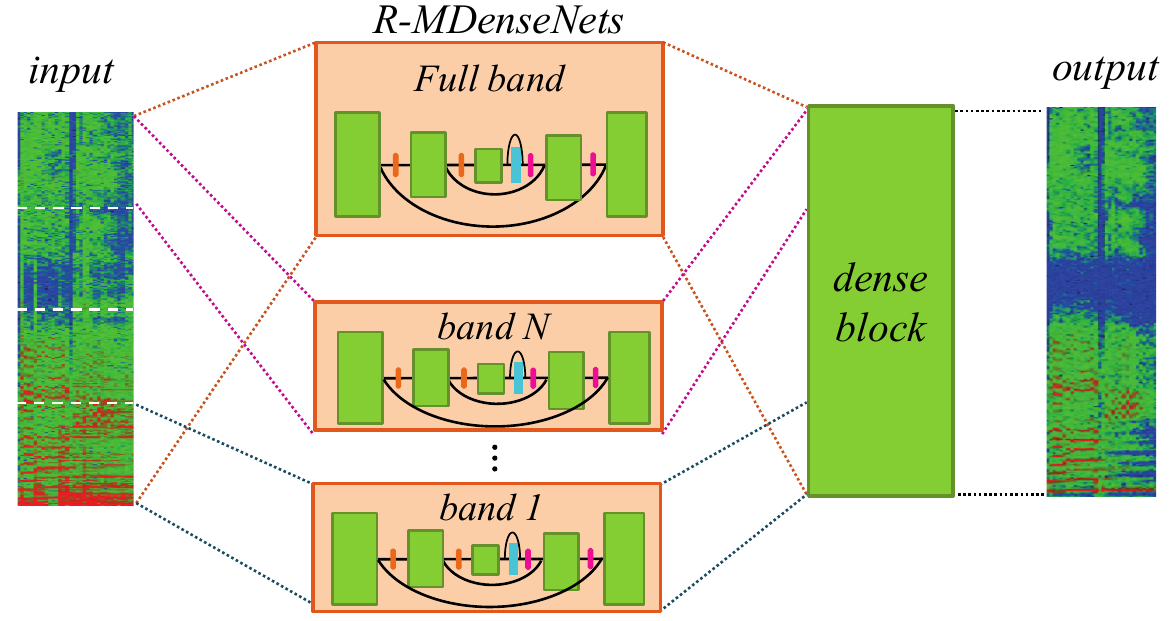}
\caption{MMDenseLSTM architecture. Outputs of MDenseLSTM dedicated to different frequency band including the full band are concatenated and the final dense block integrates features from these bands to create the final output.}
\label{fig:rmmdense}
\end{figure}

Naively inserting LSTM blocks at every scale greatly increases the model size. This is mostly due to the full connection between the input and output units of the LSTM block in the scale $s=1$. To address this problem, we propose the insertion of only a small number of LSTM blocks in the upsampling path for low scales ($s>1$). This makes it easier for LSTM blocks to capture the global structure of the input with a much smaller number of parameters. On the other hand, a CNN is advantageous for modeling fine local structures; thus placing only dense block at $s=1$ is suitable.
The multi-band structure is also beneficial for LSTM blocks since the compression from the input frequency dimension $f^s$ to $m^s$ LSTM units is relaxed or it even allows the dimension ($f^s<m^s$) to be increased while using fewer LSTM units, increasing the modeling capabilities as discussed in \cite{Zagoruyko16}. 
The entire proposed architecture is illustrated in Fig. \ref{fig:rmmdense}.
To capture the pattern that spans the bands, MDenseLSTM for the full band is also built in parallel along with the band dedicated MDenseLSTM. The outputs of the MDenseLSTMs are concatenated and integrated by the final dense block, as MMDenseNet.
\label{sec:MMDenseNet}

\subsection{Architectural details}
\label{sec:detailArch}
Details of the proposed network architecture for ASS are described in Table \ref{tab:densearch}. We split the input into three bands at $4.1$kHz and $11$kHz. The LSTM blocks are only placed at bottleneck blocks and at some blocks at $s=2$ in the upsampling path, which greatly reduces the model size. The final dense block has three layers with growth rate $k=12$. The effective context size of the architecture is 356 frames. Note that MMDenseLSTM can be applied to an input of arbitrary length since it consists of convolution and LSTM layers.

 \begin{table}[t]
    \caption{\label{tab:densearch} {\it The proposed architecture. All dense blocks are equipped with 3$\times$3 kernels with growth rate $k$.  $l$ and $m^s$ denote the number of layer and LSTM units of LSTM block, respectively. $ds$ denotes scale $s$ in the downsampling path while $us$ is that in the upsampling path.}}
    \vspace{2mm}
    \centerline{
      \small
      \tabcolsep=3px
      \begin{tabular}{ c | c | c  c  c  c  c  c  c  c  c  c} 
        \hline
        band & $k$ & scale & d1 & d2 & d3 & d4 & d5 & u4 & u3 & u2 & u1\\
        \hline
        \multirow{2}{*}{1} & \multirow{2}{*}{14} & $l$ & 5 & 5 & 5 & 5 & - & - & 5 & 5 & 5  \\
         &  &  $m^s$  & - & - & - & 128 & - & - & - & 128 & - \\
        \hline
        \multirow{2}{*}{2} & \multirow{2}{*}{4} & $l$ & 4 & 4 & 4 & 4 & - & - & 4 & 4 & 4  \\
         &  &  $m^s$  & - & - & - & 32 & - & - & - & - & - \\
        \hline
        \multirow{2}{*}{3} & \multirow{2}{*}{2} & $l$ & 1 & 1 & - & - & - & - & - & 1 & 1  \\
         &  &  $m^s$  & - & - & 8 & - & - & - & - & - & - \\
        \hline  
        \multirow{2}{*}{full} & \multirow{2}{*}{7} & $l$ & 3 & 3 & 4 & 5 & 5 & 5 & 4 & 3 & 3   \\
         &  &  $m^s$  & - & - & - & 128 & - & - & - & 128 & - \\
        \hline  
      \end{tabular}
    }
\end{table}

\section{Experiments}
\subsection{Setup}
\label{sec:setup}
We evaluated the proposed method on the DSD100 and MUSDB18 datasets, prepared for SiSEC 2016 \cite{Liutkus17} and SiSEC 2018 \cite{sisec2018}, respectively. 
MUSDB18 has 100 and 50 songs while DSD100 has 50 songs each in the {\it Dev} and {\it Test} sets.
In both datasets, a mixture and its four sources, {\it bass, drums, other} and {\it vocals}, recorded in stereo format at 44.1kHz, are available for each song.
Short-time Fourier transform magnitude frames of the mixture, windowed at 4096 samples with 75\% overlap, with data augmentation \cite{Uhlich17}  were used as inputs. 
The networks were trained to estimate the source spectrogram by minimizing the mean square error with the Adam optimizer.
For the evaluation on MUSDB18, we used the {\it museval} package 
\cite{sisec2018}, while we used the BSSEval v3 toolbox \cite{Vincent06} for the evaluation on DSD100 for a fair comparison with previously reported results. The SDR values are the median of the average SDR of each song.

\subsection{Architecture validation}
\label{sec:arcvalid}
\begin{figure}[t]
\centering
\includegraphics[width=0.84\linewidth]{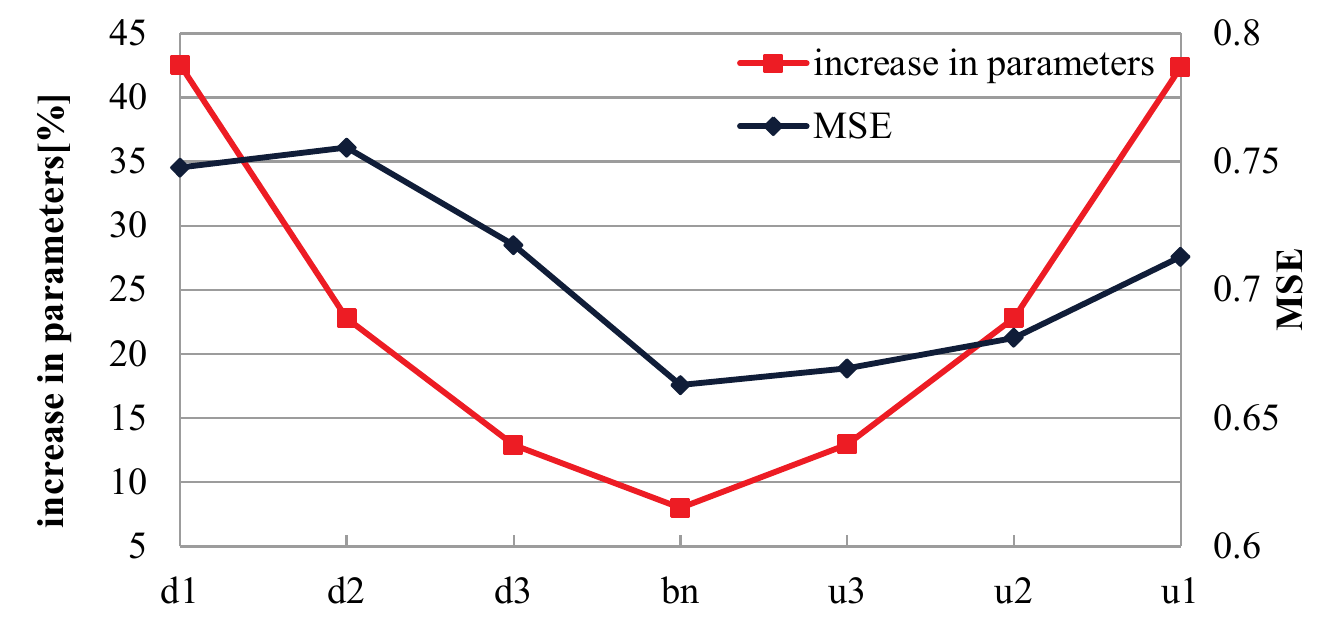}
\caption{Effect of LSTM block at different scales.}
\label{fig:lstmpos}
\end{figure}

\begin{figure}
\centering
\includegraphics[width=0.85\linewidth, height=0.2\textwidth]{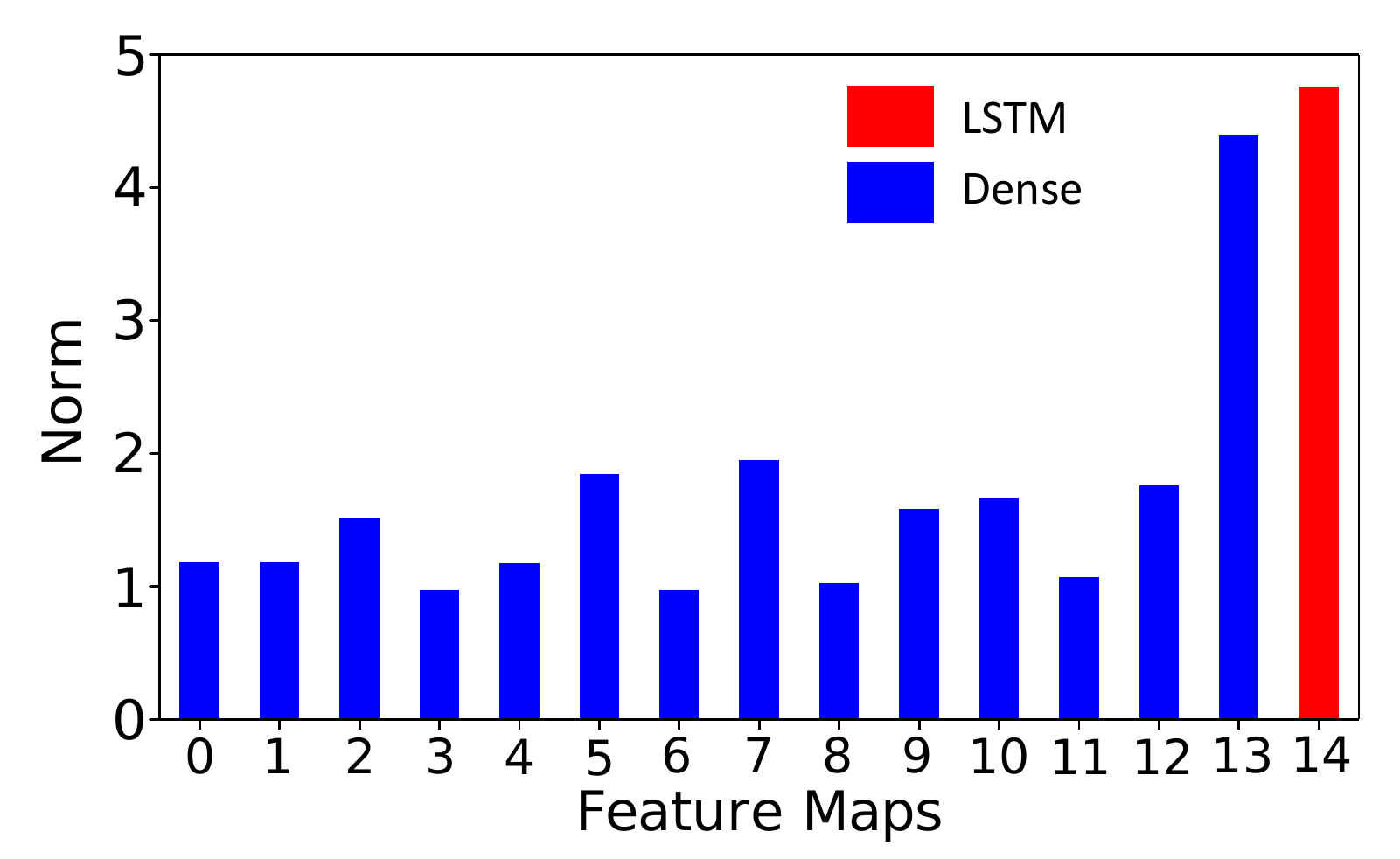}
\caption{Average $l2$ norm of feature maps.}
\label{fig:l2norm}
\end{figure}

\begin{table}[t]
\caption{\label{tab:ex1} {\it Comparison of MMDenseLSTM configurations.}}
\vspace{2mm}
\centering{
\begin{tabular}{c | c c c } 
\hline
type  		 &Sa				& Sb		&P \\
\hline\hline
SDR		 	 &{\bf 2.83}		& 2.31		& 2.47 \\
\hline
\end{tabular}
}
\end{table}  
In this section we validate the proposed architecture for the singing voice separation task on MUSDB18.\\
\textbf{Combination structure} \hspace{2mm} The SDR values obtained by the \textit{Sa-}, \textit{Sb-} and \textit{P-} type MMDenseLSTMs are tabulated in Table \ref{tab:ex1}. These results validate our claim (Sec. \ref{sec:MMDenseNet}) that the \textit{Sa} configuration performs the best because the LSTM layer can efficiently model the global modulations utilizing the local features extracted by the dense layers at this scale. Henceforth, all experiments use the \textit{Sa} configuration.\vspace{1mm}\\
\textbf{LSTM insertion scale} \hspace{3mm} The efficiency of inserting the LSTM block at lower scales was validated by comparing seven MMDenseLSTMs with a single 64 unit LSTM layer inserted at different scales in band 1 (all other LSTM layers in Table \ref{tab:densearch} are omitted). Figure \ref{fig:lstmpos} shows the percentage increase in the number of parameters compared with that of the base architecture and the mean square error (MSE) values for the seven networks. It is evident that inserting LSTM layers at low scales in the up-scaling path gives the best performance.\vspace{1mm}\\
\textbf{Contribution of dense and LSTM layers} \hspace{3mm} We further compared the $l2$ norms of the feature maps (Fig.\ref{fig:l2norm}) in the LSTM block $d4$ of band 1. It can be seen that the norm of the LSTM feature map is similar to the highest norm among the dense feature maps. Even though some dense feature maps have low norms, we confirmed that they tend to learn sparse local features.

\subsection{Comparison with state-of-the-art methods}
\begin{table}[t]
\caption{\label{tab:ex2} {\it Comparison of SDR on DSD100 dataset.}}
\vspace{2mm}
\centering{
\resizebox{\linewidth}{!}{
\begin{tabular}{ c | c c c c c} 
\hline
\multicolumn{1}{c|}{} & \multicolumn{5}{c}{SDR in dB}\\
Method      &	Bass	&	Drums	& Other & Vocals & Acco.	\\
\hline\hline
DeepNMF \cite{Roux15}	&	1.88	& 2.11 & 2.64 & 2.75 &  8.90 \\
NUG \cite{Nugraha15}\	&	2.72	& 3.89 & 3.18 & 4.55 & 10.29 \\
BLSTM \cite{Uhlich17}\	&	2.89	& 4.00 & 3.24 & 4.86 & 11.26 \\
BLEND \cite{Uhlich17}\  &	2.98    & 4.13 & 3.52 & 5.23 & 11.70 \\
MMDenseNet \cite{Takahashi17MMDense}\ 	&	{\bf 3.91}	& 5.37 & 3.81 & 6.00 & 12.10 \\
{\bf MMDenseLSTM} 	&	3.73	& {\bf 5.46} & {\bf 4.33} & {\bf 6.31} & {\bf 12.73}\\

\hline
\end{tabular}
}}
\end{table} 

\begin{table}[t]
\caption{\label{tab:ex2_1} {\it Comparison of SDR on MUSDB18 dataset.}}
\vspace{2mm}
\centering{
\resizebox{\linewidth}{!}{
\begin{tabular}{ c | c | c c c c c} 
\hline
\multicolumn{1}{c|}{} & \multicolumn{1}{c|}{\#params} & \multicolumn{5}{c}{SDR in dB}\\
Method  & [$\times10^6$]    &	Bass	&	Drums	& Other & Vocals & Acco.	\\
\hline\hline
IBM \	& - &	5.30	& 6.87 & 6.42 & 7.50 & 10.83 \\
\hline
BLSTM \cite{Uhlich17}\	& 30.03 &	3.99	& 5.28 & 4.06 & 3.43 & 14.51 \\
MMDenseNet \cite{Takahashi17MMDense}\ & 0.33	&	{\bf 5.19}	& 6.27 & 4.64 & 3.87 & 15.41 \\
BLEND2 \ & 30.36	&	4.72	& 6.25 & 4.75 & 4.33 & 16.04 \\
{\bf MMDenseLSTM} 	& 1.22 &	{\bf 5.19}	& {\bf 6.62} & {\bf 4.93} & {\bf 4.94} & {\bf 16.40}\\

\hline
\end{tabular}
}
}
\end{table} 
We next compared the proposed method with five state-of-the-art methods, DeepNMF \cite{Roux15}, NUG \cite{Nugraha15}, BLSTM \cite{Uhlich17}, BLEND \cite{Uhlich17} and MMDenseNet \cite{Takahashi17MMDense} on DSD100. The task was to separate the four sources and accompaniment, which is the residual of the vocal extraction, from the mixture. Here, the multichannel Wiener filter was applied to MMDenseLSTM outputs as in \cite{Uhlich17,Takahashi17MMDense}.
Table \ref{tab:ex2} shows that the proposed method improves SDRs by an average of $0.2$dB compared with MMDenseNet, showing that the MMDenseLSTM architecture further improves the performance for most sources.

To further improve the capability of music source separation and utilize the full modeling capability of MMDenseLSTM, we next trained models with the MUSDB {\it dev} set and an internal dataset comprising 800 songs resulting in a $14$ times larger than the DSD100 {\it dev} set.
The proposed method was compared with BLSTM \cite{Uhlich17}, MMDenseNet \cite{Takahashi17MMDense} and a blend of these two systems (BLEND2) as in \cite{Uhlich17}. All baseline networks were trained with the same training set, namely 900 songs. For a fair comparison with MMDenseNet, we configured it with the same base architecture as in Table \ref{tab:densearch}, with an extra layer in the {\it dense blocks}, corresponding to the {\it LSTM block} in our proposed method. 
We also included the IBM as an upper baseline since it uses oracle separation.
Table \ref{tab:ex2_1} 
shows the result of this experiment. We obtained average improvements of 0.43dB over MMDenseNet and 0.41dB over BLEND2, achieving state-of-the-art results in SiSEC2018 \cite{sisec2018}. The proposed method even outperformed the IBM for {\it accompaniment}. Table \ref{tab:ex2_1} also shows that MMDenseLSTM can efficiently utilize the sequence modeling capability of LSTMs in conjunction with MMDenseNet, having 24 times fewer parameters than the naive combination of BLSTM and MMDenseNet.

\section{Conclusion}
\label{sec:concl}
We proposed an efficient way to combine DenseNet and LSTM to improve ASS performance. The proposed MMDenseLSTM achieves state-of-the-art results on DSD100 and MUSDB18 datasets. MMDenseLSTM outperforms a naive combination of BSLTM and MMDenseNet despite having much fewer parameters, and even outperforms an IBM for a singing voice separation task when the networks were trained with 900 songs. The improvement over MMDenseNet is less for {\it bass}, which will be further investigated in future.

\bibliographystyle{IEEEbib}
\bibliography{bss,CV,other}

\end{document}